\documentclass[a4paper,11pt]{article}

\usepackage{pos}
\usepackage{siunitx}

\title{Probing ultra-high-energy neutrinos with the IceCube-Gen2 in-ice radio array}

\ShortTitle{IceCube-Gen2 radio}

\author{The IceCube-Gen2 Collaboration \\{\normalsize \normalfont(a complete list of authors can be found at the end of the proceedings)}\\}

\emailAdd{christian.glaser@physics.uu.se}

\abstract{

The next generation neutrino telescope, IceCube-Gen2, will be sensitive to the astrophysical and cosmogenic flux of neutrinos across a broad energy range, from the TeV to the EeV scale. The planned design includes 8 cubic kilometers of ice instrumented with approximately 10,000 optical sensors, a surface array, and a radio array of antennas embedded in the ice laid out sparsely over \SI{500}{km^2}. The radio array provides sensitivity to ultra-high energy neutrinos using independent radio stations that can trigger on Askaryan emission from neutrino interactions in the ice. In this contribution, we present the design for the radio array along with its planned implementation, which is expected to increase sensitivity to neutrinos with energies beyond \SI{100}{PeV} by at least an order of magnitude over existing arrays. Furthermore, we will quantify the expected science output by presenting measurement forecasts for the main science cases of diffuse flux and point source discovery, as well as cross-section and flavor measurements.
\vspace{4mm}

{\bfseries Corresponding authors:}
Christian Glaser$^{1*}$\\
{$^{1}$ \itshape Uppsala University, Sweden}\\[4mm]
$^*$ Presenter
}

\ConferenceLogo{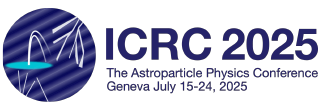}

\FullConference{39th International Cosmic Ray Conference (ICRC2025)\\
 15–24 July 2025\\
Geneva, Switzerland\\}

\begin{document}

\maketitle

\section{Introduction}

With the first detection of high-energy cosmic neutrinos of extraterrestrial origin in 2013~\cite{IceCube:2013low}, the IceCube Neutrino Observatory opened a new window to some of the most extreme regions of our universe.
With IceCube-Gen2 \cite{IceCube-Gen2-TDR}, we propose a next-generation detector that will increase the neutrino collection rate by an order of magnitude and increase the energy reach towards EeV energies. IceCube-Gen2 will be a unique wide-band neutrino observatory (MeV--EeV) that employs two complementary detection technologies for neutrinos, optical and radio, combined with a surface detector array for cosmic-ray air showers. The IceCube-Gen2 project completed its Technical Design Report \cite{IceCube-Gen2-TDR} and it taking the next step towards a preliminary design review.

Here, we describe the radio component of IceCube-Gen2, which increases the neutrino sensitivity towards EeV energies. Neutrino interactions initiate large particle cascades that generate short nanoseconds-long radio flashes through the Askaryan effect. 
Expanding the energy range to EeV (=$10^{18}$\,eV) energies requires the instrumentation of much larger volumes to cope with the decreasing neutrino flux, which can be achieved cost-efficiently by a sparse array of relatively shallow radio detector stations due to the long kilometer-scale attenuation length of radio signals in ice \cite{Barwick:2022vqt}. 

Due to the extremely low neutrino flux at energies above \SI{10}{PeV}, no neutrino has yet been detected using the radio technique. However, several experiments have shown the feasibility of this detection method and its potential. It builds on the experience of previous radio neutrinos detectors, like the pioneering RICE and ANITA experiments, which informed the development of prototype in-ice arrays such as ARIANNA and ARA experiments (see \cite{Barwick:2022vqt} and references therein). 
Currently, the Radio Neutrino Observatory in Greenland (RNO-G), which has been under construction since 2021 \cite{RNO-G:2020rmc}, is operational with 8 out of the 35 planned stations \cite{RNOG-ICRC25-overview}. RNO-G is an order of magnitude smaller than IceCube-Gen2, but when commissioned, it will be significantly larger than the pilot arrays ARA and ARIANNA, and has the potential to discover the first UHE neutrino (see Fig.~\ref{fig:diffuse_sensitivity}). The construction of RNO-G will continue over the following years and also serve as a development site for the radio array of IceCube-Gen2.

\begin{figure}
    \centering
    \includegraphics[width=0.7\linewidth]{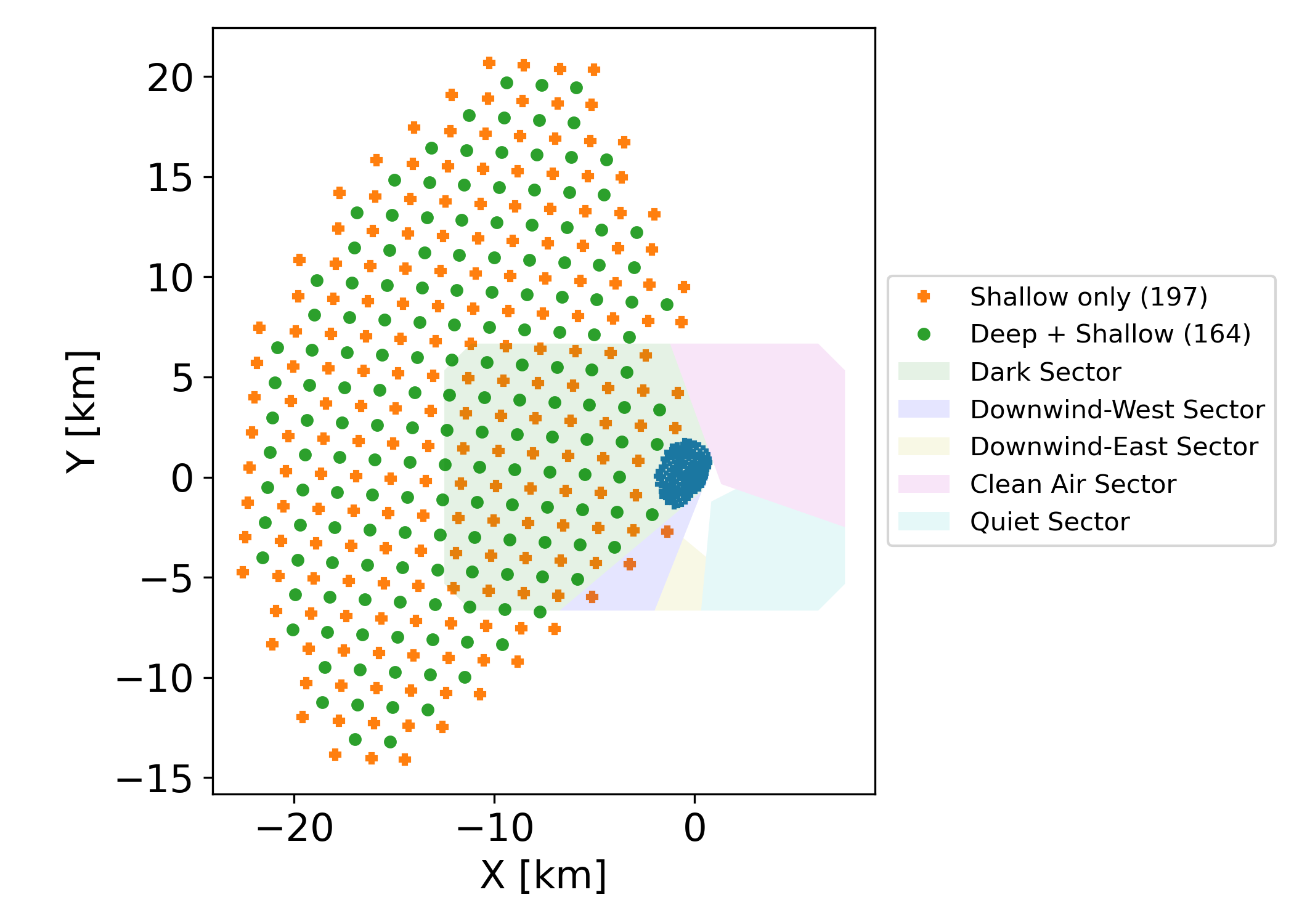}
    \caption{The locations of the radio array stations. The shallow-only stations are shown in orange crosses, while the hybrid stations are shown as green circles.  Figure and caption reproduced from \cite{IceCube-Gen2-TDR}.}
    \label{fig:radio_array_layout}
\end{figure}

\section{IceCube-Gen2 radio baseline design}

The radio array of IceCube-Gen2 targets the discovery and characterization of the neutrino flux above 10~PeV. 
Radio emission can be observed from showers from neutral-current interactions (all flavor) as well as charged-current interactions  ($\nu_e$) or catastrophic energy losses of secondary leptons ($\nu_{\mu}, \nu_{\tau}$), which allows the radio detector to obtain sensitivity to the neutrino flavor \cite{Garcia-Fernandez:2020dhb,Coleman:2024scd}.

The reference design of the radio array \cite{IceCube-Gen2-TDR} consists of 361 individual stations (see Fig.~\ref{fig:radio_array_layout}), each with multiple antennas sensitive to different electric field polarizations. 
By design, only a small fraction of neutrino events are detectable in more than one station, such that the total effective volume is (approximately) the linear sum of the per-station effective volume (see Fig.~\ref{fig:radio_array_layout}). 

\begin{figure}[t]
    \centering
    \includegraphics[width=0.7\linewidth]{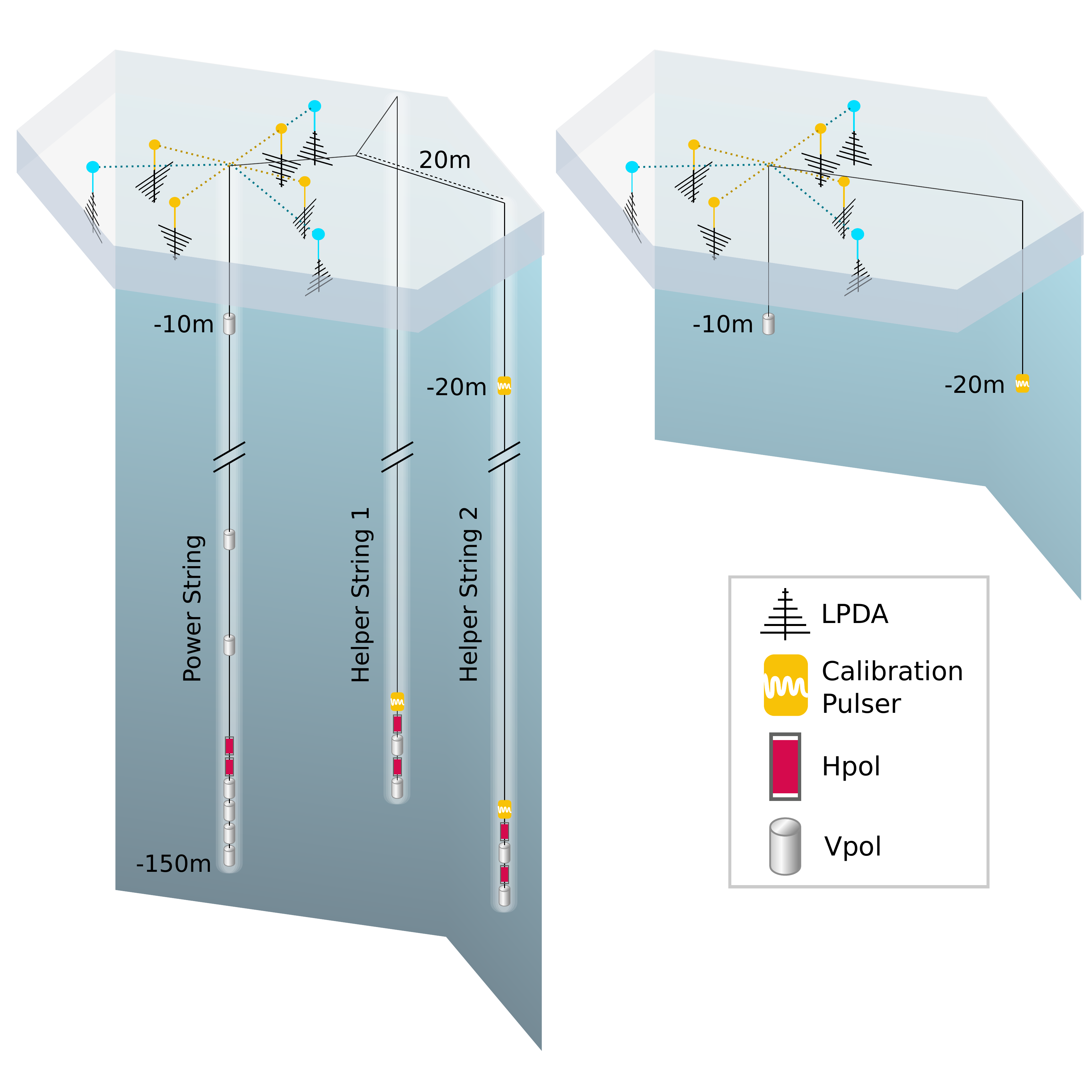}
    \caption{Illustration of the hybrid station layouts (left) and the shallow-only station layout (right) of the reference design. Figure and caption reproduced from \cite{IceCube-Gen2-TDR}.}
    \label{fig:stations}
\end{figure}

The array will combine two different types of stations. 197 so-called \textit{shallow stations} comprise 8\,channels/antennas, mostly log-periodic dipole antennas, close to the surface (see Fig.~\ref{fig:stations}~right). The other station design is the \textit{hybrid station}, which has the same shallow component, but adds a phased array of antennas at \SI{150}{m} depth with additional antennas on the same string and two additional strings for event reconstruction and background rejection (see Fig.~\ref{fig:stations}~left). 164 of these hybrid stations with 24 channels/antennas will be built. Due to the downward bending of radio signals in the firn, deeper antennas offer higher per-station neutrino sensitivity. However, the increased deployment cost offsets this gain, resulting in similar overall sensitivity per unit cost between shallow and deep station designs.

\begin{figure}
    \centering
    \includegraphics[width=0.55\linewidth]{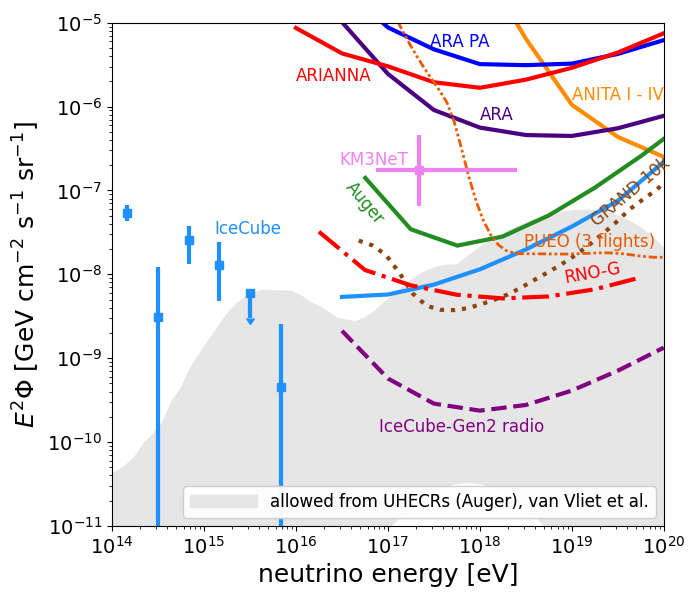}
    \includegraphics[width=0.44\linewidth]{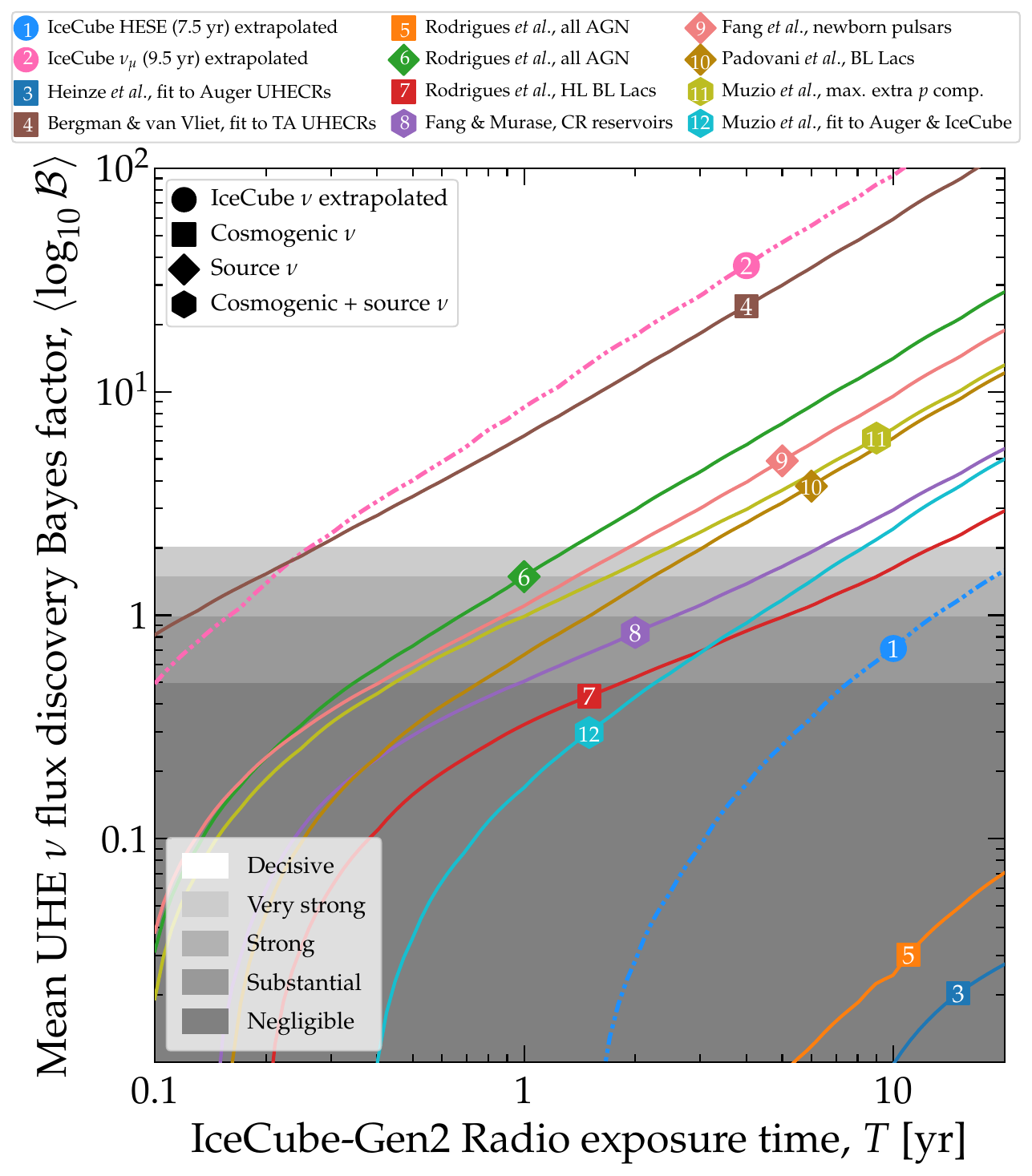}
    \caption{(left) Expected sensitivity of the IceCube-Gen2 in-ice radio detector at the highest energies in comparison to the allowed region of GZK neutrinos based on Auger measurements \cite{Vliet2019}, measurements of IceCube \cite{IceCube-Gen2-TDR} and KM3Net \cite{KM3NeT:2025npi}, existing upper limits \cite{Anker:ARIANNAlimit2019, ARA:2019wcf, ARA:2022rwq, AugerNuLimit2019, IceCube:2025ezc}, and expected sensitivities of RNO-G (under construction in Greenland) \cite{RNO-G:2020rmc}, the PUEO balloon mission (to be launched in 2026) \cite{PUEO:2020bnn}, and the proposed GRAND array with 10,000 detector stations \cite{GRAND:2018iaj}. (right) Discovery potential of several diffuse ultra-high-energy (UHE) neutrino flux models (see \cite{Valera:2022wmu} and references therein for details). Figure from \cite{Valera:2022wmu}.} 
    \label{fig:diffuse_sensitivity}
\end{figure}

The stations are spaced on a square grid of \SI{1.24}{km} spacing, and alternate between shallow and hybrid stations. With this spacing, at least 10\% of the neutrino events will be detected in multiple stations, comprising a sample of events with significantly enhanced reconstruction.  The overall dimensions are chosen to be within the Dark Sector at the South Pole to minimize radio interference and comply with agreements from other experiments. The expected sensitivity to a diffuse flux of neutrinos of the radio component of IceCube-Gen2 is shown in Fig.~\ref{fig:diffuse_sensitivity} alongside existing measurements and limits and a selection of expected sensitivities of other ongoing projects. 

The advantage of a hybrid array lies in the complementarity between the two station types, which significantly reduces systematic uncertainties and experimental risks, thus making it ideal for discovering and characterizing the UHE neutrino flux. 
The two components rely on independent triggering and background rejection techniques, allowing for cross-validation of signal candidates. Neutrino-induced events detected by both station types would greatly strengthen confidence in a genuine UHE neutrino detection. Furthermore, the differing reconstruction approaches and signal propagation paths through the ice provide an opportunity to constrain and validate reconstruction systematics experimentally and to mitigate uncertainties related to ice properties. 
Deployment risks are further mitigated as shallow stations are simpler and less expensive to install since they do not require deep drilling. While achieving comparable sensitivity requires a larger number of shallow stations, their ease of deployment allows for flexible scaling and adaptive scheduling, making the array more resilient to delays.
Therefore, a hybrid design with similar neutrino sensitivity in both components enhances the experiment's robustness. It reduces deployment risk, decreases systematic uncertainties, and increases confidence in detecting and characterizing the UHE neutrino flux.

\section{Forecast of main science cases}
In this section, we briefly summarize IceCube-Gen2 radio's capabilities in the four main science cases of discovering and characterizing the ultra-high-energy neutrino flux, discovering sources, measuring the neutrino-nucleon cross-section, and determining the neutrino flavor composition.

In \cite{Valera:2022wmu}, a detailed forecast was presented for the discovery potential of a diffuse flux of ultra-high-energy (UHE) neutrinos with IceCube-Gen2 radio. The study considered a broad set of benchmark flux models, including cosmogenic and astrophysical scenarios as well as extrapolations of the IceCube TeV–PeV flux to higher energies. The analysis incorporated realistic modeling of neutrino propagation through the Earth, expected event rates, detector response, and background contributions, while accounting for key theoretical and experimental uncertainties, such as those associated with the UHE neutrino-nucleon cross section. A Bayesian statistical framework was employed to evaluate discovery prospects, showing that most benchmark models could be detected within a few years of operation, and some within months (see Fig.~\ref{fig:diffuse_sensitivity} right for details). The high-energy tail of the IceCube neutrino flux was identified as a significant background, underscoring the importance of constraining its spectral cutoff. Additionally, the study found that a successful detection could enable discrimination between most flux models within a decade.

In \cite{Fiorillo:2022ijt}, forecasts were presented for discovering UHE neutrinos above \SI{100}{PeV} through searches for UHE neutrino multiplets with IceCube-Gen2 radio. The study demonstrated that such a discovery could be achieved within ten years of operation and would impose strong constraints on the population of UHE cosmic-ray sources. The analysis included state-of-the-art modeling of neutrino propagation, radio detection, and relevant backgrounds. For steady sources, the detection of even a single multiplet would disfavor most candidate classes as individually dominant, while a non-detection would rule out only the brightest and rarest sources. For transient sources, non-detection would disfavor classes with total energy outputs above $10^{53}$~erg. The results were found to be robust against uncertainties in background levels, detector volume, and energy resolution, but sensitive to angular resolution, motivating a target zenith-angle resolution of approximately \SI{2}{\degree}.

\begin{figure}
    \centering
    \includegraphics[width=0.8\linewidth]{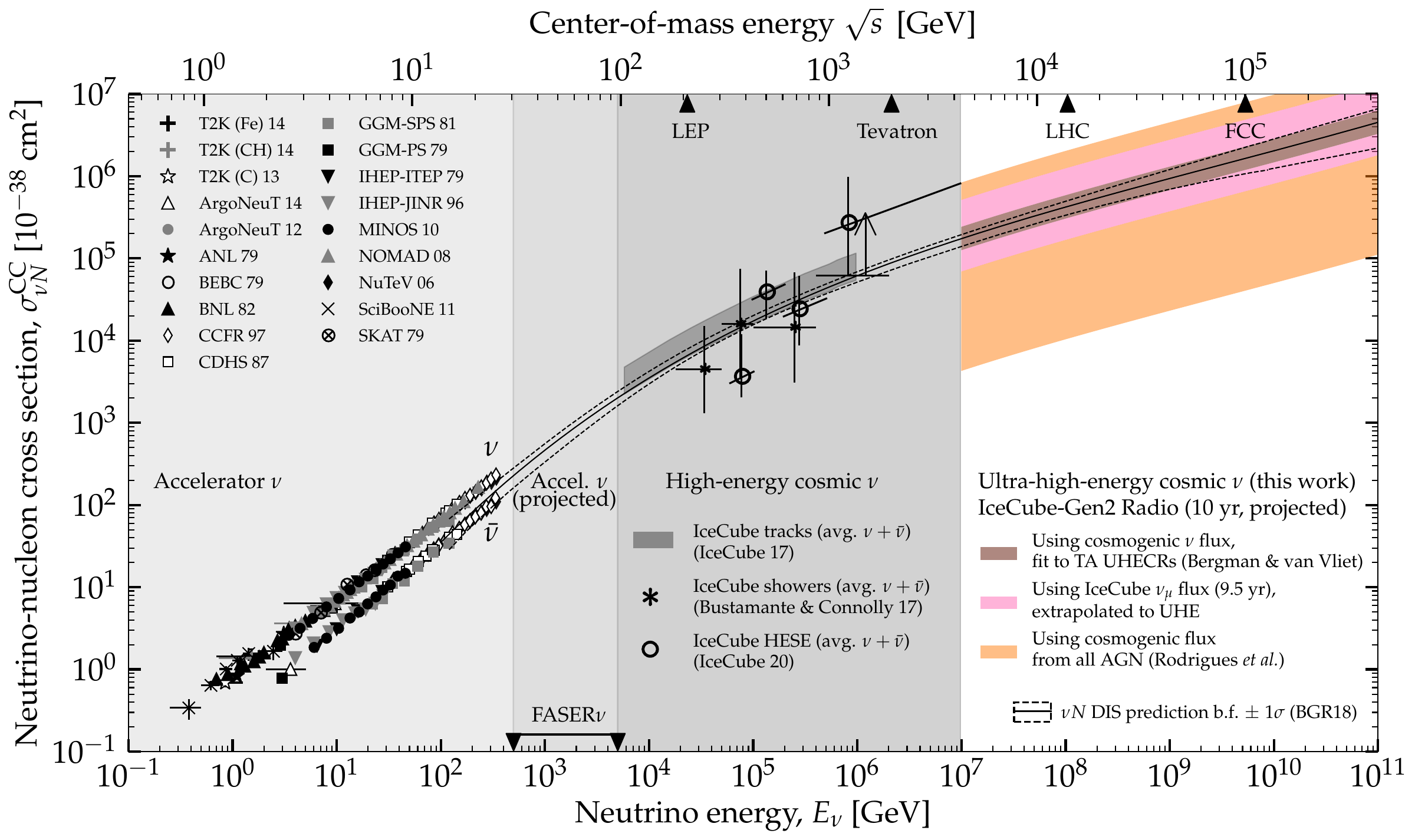}
    \caption{Neutrino-nucleon ($\nu N$) charged-current (CC) cross section, measurements and predictions. See \cite{Valera:2022ylt} and references therein for details. The radio detector of IceCube-Gen2 will provide sensitivity above 100~PeV where no measurement exists presently.  Shown are the 10-year forecasts for three representative flux scenarios. For comparison, the state-of-the-art BGR18 calculation of the $\nu N$ deep-inelastic-scattering cross section is shown. Figure and caption adapted from \cite{Valera:2022ylt}.}
    \label{fig:cross-section}
\end{figure}

In \cite{Valera:2022ylt}, the first detailed forecasts were presented for measuring the ultra-high-energy (UHE) neutrino-nucleon ($\nu N$) cross section using IceCube-Gen2 radio. The sensitivity to the cross section arises from the directional dependence of UHE neutrinos arriving at the detector after propagation through the Earth, particularly from Earth-skimming trajectories where interaction probabilities are significant. The analysis employed the BGR18 deep-inelastic-scattering (DIS) calculation as the baseline cross section model and incorporated a broad range of UHE neutrino flux predictions, including cosmogenic and astrophysical sources as well as extrapolations of the IceCube TeV–PeV flux. A Bayesian statistical framework was used to extract the $\nu N$ cross section jointly with the flux normalization, incorporating expected event statistics and detector resolution. The study found that, assuming the detection of several tens of neutrino-induced showers, the $\nu N$ cross section could be measured to within approximately 50\% of the standard-model prediction within 10 years, and possibly within 5 years under optimistic assumptions (see Fig.~\ref{fig:cross-section}). The dominant systematic uncertainty arises from the unknown UHE neutrino flux, but the projected precision is sufficient to test standard QCD predictions and probe potential beyond-standard-model effects such as non-linear QCD dynamics, color-glass condensates, and sphaleron-induced processes.

In \cite{Coleman:2024scd}, a method was presented for measuring the flavor composition of a diffuse flux of UHE neutrinos in upcoming in-ice radio-detection neutrino telescopes, focusing on IceCube-Gen2 radio.  Sensitivity to the electron neutrino component arises from the identification of charged-current $\nu_e$ interactions, aided by a neural network trained to recognize signatures of the Landau-Pomeranchuk-Migdal effect. Sensitivity to $\nu_\mu + \nu_\tau$ is achieved via detection of secondary muon and tau interactions, allowing for spatially separated Askaryan signals within the detector array. Assuming a high UHE neutrino flux yielding approximately 180 detected events in ten years, the study found that IceCube-Gen2 would be capable of distinguishing between benchmark production scenarios—pion decay, muon-damped decay, and neutron decay—at greater than 68\% confidence level. The results indicate that flavor composition measurements could provide valuable constraints on production mechanisms and offer sensitivity to possible new physics at ultra-high energies.

\begin{figure}
    \centering
    \includegraphics[width=0.9\linewidth]{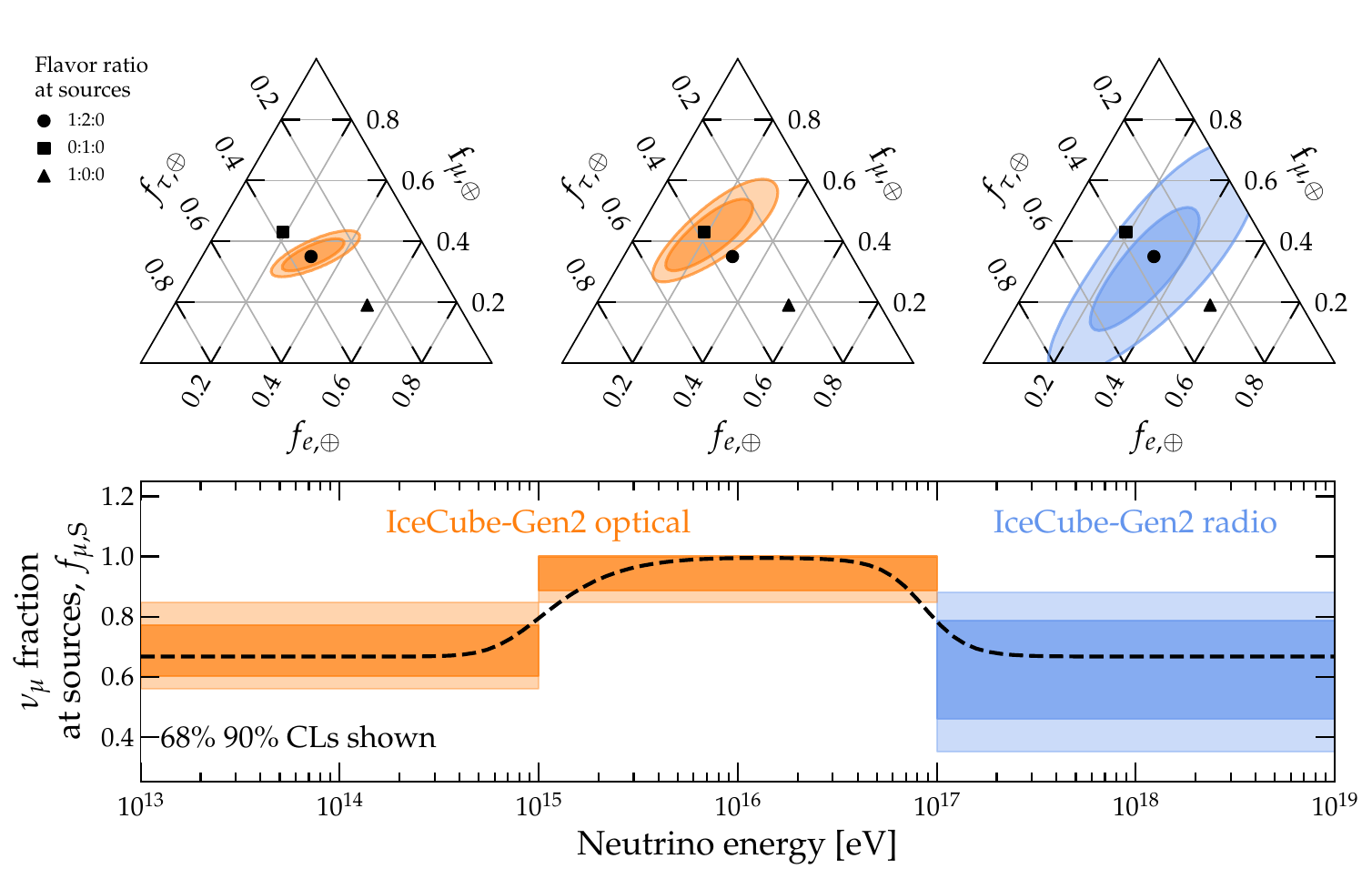}
    \caption{IceCube-Gen2 capabilities to measure the neutrino flavor composition from TeV to EeV. The first two bins are measured with the optical detector of IceCube-Gen2~\cite{IceCube-Gen2-TDR} and the last bin with the radio detector \cite{Coleman:2024scd}. The astrophysical neutrino sources are assumed to be dominated by pion decay at the lowest energies, $\left(\frac{1}{3},\frac{2}{3},0\right)_{\rm S}$, transition to muon-damped production at intermediate energies, $\left(0,1,0\right)_{\rm S}$, and finally return to the pion decay composition expected from cosmogenic neutrino production. \textit{Top:} Uncertainty in the measurement of the flavor composition at Earth, $f_{\alpha, \oplus}$.  \textit{Bottom:} Assumed model for the evolution of $\nu_\mu$ content at the sources, $f_{\mu, {\rm S}}$, and associated uncertainty in its inferred value from measurements. Figure and caption reproduced from \cite{Coleman:2024scd}.}
    \label{fig:flavor}
\end{figure}

Furthermore, the unique capability of IceCube-Gen2 to combine its optical and radio components enables a continuous measurement of the neutrino flavor composition across six orders of magnitude in energy, from TeV to EeV (see Fig.~\ref{fig:flavor}). This combined approach allows IceCube-Gen2 to probe energy-dependent transitions in the flavor composition predicted by astrophysical source models, such as the shift from pion decay to muon-damped and eventually to kaon-damped regimes, and to distinguish between astrophysical and cosmogenic neutrino contributions at the highest energies.

\section{Outlook}
While the reference design \cite{IceCube-Gen2-TDR} fulfills the science objectives, work is ongoing to further enhance the performance of the IceCube-Gen2 radio detector \cite{Glaser:2023udy}. Two factors are central to increasing its scientific output: the UHE neutrino detection rate and the precision of energy and directional reconstruction. 
The detection rate can be increased by replacing the current threshold-based triggers with neural network-based algorithms, which could double the detection rate by better distinguishing signal waveforms from thermal noise -- the dominant background at the trigger level. Prototype implementations on existing hardware have shown promising results \cite{ Arianna:2021vcx, RNO-G:2023oxb}, and a new DAQ system with onboard AI processing is being developed to support this which will be tested in RNO-G.

In parallel, work is underway to optimize the detector station geometry specifically for accurate energy and directional reconstruction. A differentiable simulation and reconstruction pipeline is being developed to enable full end-to-end optimization via gradient descent. 
For optimistic assumptions, it was estimated that both trigger and station layout improvements would expedite the discovery of UHE neutrino fluxes by up to a factor of five, see sources from deeper in our Universe increasing the observable volume by a factor of three, and measure the neutrino-nucleon cross-section at EeV energies with 3x smaller uncertainty \cite{Glaser:2023udy}. 

\section{Summary}

The planned in-ice radio array of IceCube-Gen2 will significantly extend the sensitivity of neutrino telescopes into the ultra-high-energy regime, enabling the detection of astrophysical and cosmogenic neutrinos above \SI{100}{PeV}. By combining the radio array with the optical and surface components, IceCube-Gen2 offers a uniquely broad energy reach from TeV to EeV scales. The hybrid design of the radio array mitigates systematic uncertainties, reduces deployment risks, and enhances robustness in event reconstruction and signal identification. Measurement forecasts demonstrate strong discovery potential for diffuse fluxes, point sources, and UHE neutrino interactions, including flavor and cross-section measurements. With ongoing developments in triggering, reconstruction, and array optimization, IceCube-Gen2 is poised to deliver transformative contributions to astroparticle physics and the study of the high-energy universe.

\bibliographystyle{ICRC}
\setlength{\bibsep}{2.5pt}
\bibliography{references}

\clearpage

\section*{Full Author List: IceCube-Gen2 Collaboration}

\scriptsize
\noindent
R. Abbasi$^{16}$,
M. Ackermann$^{76}$,
J. Adams$^{21}$,
S. K. Agarwalla$^{46,\: {\rm a}}$,
J. A. Aguilar$^{10}$,
M. Ahlers$^{25}$,
J.M. Alameddine$^{26}$,
S. Ali$^{39}$,
N. M. Amin$^{52}$,
K. Andeen$^{49}$
% ...
\\
\\
% ...
$^{10}$ Universit{\'e} Libre de Bruxelles, Science Faculty CP230, B-1050 Brussels, Belgium \\
% ...
$^{16}$ Department of Physics, Loyola University Chicago, Chicago, IL 60660, USA \\
% ...
$^{21}$ Dept. of Physics and Astronomy, University of Canterbury, Private Bag 4800, Christchurch, New Zealand \\
% ...
$^{25}$ Niels Bohr Institute, University of Copenhagen, DK-2100 Copenhagen, Denmark \\
$^{26}$ Dept. of Physics, TU Dortmund University, D-44221 Dortmund, Germany \\
% ...
$^{39}$ Dept. of Physics and Astronomy, University of Kansas, Lawrence, KS 66045, USA \\
% ...
$^{46}$ Dept. of Physics and Wisconsin IceCube Particle Astrophysics Center, University of Wisconsin{\textemdash}Madison, Madison, WI 53706, USA \\
% ...
$^{49}$ Department of Physics, Marquette University, Milwaukee, WI 53201, USA \\
% ...
$^{52}$ Bartol Research Institute and Dept. of Physics and Astronomy, University of Delaware, Newark, DE 19716, USA \\
% ...
$^{76}$ Deutsches Elektronen-Synchrotron DESY, Platanenallee 6, D-15738 Zeuthen, Germany \\
$^{\rm a}$ also at Institute of Physics, Sachivalaya Marg, Sainik School Post, Bhubaneswar 751005, India \\
% ...

\subsection*{Acknowledgments}

\noindent
The authors gratefully acknowledge the support from the following agencies and institutions:
USA {\textendash} U.S. National Science Foundation-Office of Polar Programs,
U.S. National Science Foundation-Physics Division,
U.S. National Science Foundation-EPSCoR,
U.S. National Science Foundation-Office of Advanced Cyberinfrastructure,
Wisconsin Alumni Research Foundation,
Center for High Throughput Computing (CHTC) at the University of Wisconsin{\textendash}Madison,
Open Science Grid (OSG),
Partnership to Advance Throughput Computing (PATh),
Advanced Cyberinfrastructure Coordination Ecosystem: Services {\&} Support (ACCESS),
Frontera and Ranch computing project at the Texas Advanced Computing Center,
U.S. Department of Energy-National Energy Research Scientific Computing Center,
Particle astrophysics research computing center at the University of Maryland,
Institute for Cyber-Enabled Research at Michigan State University,
Astroparticle physics computational facility at Marquette University,
NVIDIA Corporation,
and Google Cloud Platform;
Belgium {\textendash} Funds for Scientific Research (FRS-FNRS and FWO),
FWO Odysseus and Big Science programmes,
and Belgian Federal Science Policy Office (Belspo);
Germany {\textendash} Bundesministerium f{\"u}r Forschung, Technologie und Raumfahrt (BMFTR),
Deutsche Forschungsgemeinschaft (DFG),
Helmholtz Alliance for Astroparticle Physics (HAP),
Initiative and Networking Fund of the Helmholtz Association,
Deutsches Elektronen Synchrotron (DESY),
and High Performance Computing cluster of the RWTH Aachen;
Sweden {\textendash} Swedish Research Council,
Swedish Polar Research Secretariat,
Swedish National Infrastructure for Computing (SNIC),
and Knut and Alice Wallenberg Foundation;
European Union {\textendash} EGI Advanced Computing for research;
Australia {\textendash} Australian Research Council;
Canada {\textendash} Natural Sciences and Engineering Research Council of Canada,
Calcul Qu{\'e}bec, Compute Ontario, Canada Foundation for Innovation, WestGrid, and Digital Research Alliance of Canada;
Denmark {\textendash} Villum Fonden, Carlsberg Foundation, and European Commission;
New Zealand {\textendash} Marsden Fund;
Japan {\textendash} Japan Society for Promotion of Science (JSPS)
and Institute for Global Prominent Research (IGPR) of Chiba University;
Korea {\textendash} National Research Foundation of Korea (NRF);
Switzerland {\textendash} Swiss National Science Foundation (SNSF).

\end{document}